\documentclass[seceq]{ptptex}
\usepackage{wrapft}
\usepackage{graphicx}

\newcommand{\wtilde}[1]{\widetilde{#1}} 

\def\beq{\begin{eqnarray}}
\def\eeq{\end{eqnarray}}
\def\bsub{\begin{subequations}}
\def\esub{\end{subequations}}
\def\b{\begin{equation}}

\title{
On the Color-Singlet States in Many-Quark Model with the $su(4)$-Algebraic Structure. II
}
\subtitle{
Determination of Ground-State Energies
}    

\author{
Yasuhiko {\sc Tsue},$^{1}$ 
Constan\c{c}a {\sc Provid\^encia},$^{2}$ 
Jo\~ao da {\sc Provid\^encia}$^{2}$ and 
Masatoshi {\sc Yamamura}$^{3}$  
}

\inst{
$^{1}$Physics Division, Faculty of Science, Kochi University, Kochi 780-8520, 
Japan\\
$^{2}$Departamento de F\'{i}sica, Universidade de Coimbra, 3004-516 Coimbra, 
Portugal\\
$^{3}$Department of Pure and Applied Physics, 
Faculty of Engineering Science,\\
Kansai University, Suita 564-8680, Japan
\\

}

\recdate{
\today
}

\abst{
Ground-state energies are investigated in a many-quark model with pairing interactions, 
which has the $su(4)$-algebraic structure. 
Exact eigenstates in the boson realization method are constructed by imposing a color-singlet condition 
developed in the previous paper. 
An interaction term breaking the $su(4)$-dynamical symmetry plays an important role to determine 
the ground state. 
As a result, a quark-pairing state or a quark-triplet state as a nucleon is realized with a certain value of 
a variable which is regarded as an order parameter. 
In addition to the parameter regions in which these ground states are realized, 
it is shown that there are two transition regions between 
the quark-pairing and the quark-triplet states with different values of the order parameter. 
}

\begin{document}

\maketitle

\section{Introduction}

The Bonn quark model is an interesting model, which was first introduced with a purpose of 
describing the nucleon and the $\Delta$-resonance as quark-triplet states.\cite{1}
The original Bonn quark model has the following Hamiltonian
\beq\label{1-1}
{\wtilde H}=-\sum_m\sum_{m'}(c_{2m}^*c_{3{\wtilde m}}^*c_{3{\wtilde m}'}c_{2m'}
+c_{3m}^*c_{1{\wtilde m}}^*c_{1{\wtilde m}'}c_{3m'}+c_{1m}^*c_{2{\wtilde m}}^*c_{2{\wtilde m}'}c_{1m'})\ , 
\eeq
where $c_{im}^*$ and $c_{im}$ are quark creation and annihilation operators with 
color $i$ and the angular momentum quantum number of the single quark level, $m=-j_s, -j_s+1 \ ,\cdots ,j_s$. 
Here, $c_{i{\wtilde m}}^*=(-1)^{j_s-m}c_{i-m}^*$. 
This Hamiltonian represents the quark-pairing interaction which, in general, leads to 
the color instability. 
Namely, it is possible that a color superconducting state is realized. 
It has been remarked that the original Bonn quark model has a dynamical $su(4)$-symmetry.\cite{A} 
Thus, a $su(4)$-symmetry breaking term is introduced in which the color $su(3)$-symmetry 
is retained. 
Namely, a $su(4)$-symmetry breaking interaction proportional to the $su(3)$-Casimir operator, 
${\wtilde {\mib Q}}^2$, 
is introduced, which represents a particle-hole-type interaction in terms of the 
quark shell model: 
\beq\label{1-2}
{\wtilde H}_m={\wtilde H}+\chi{\wtilde {\mib Q}}^2 \ .
\eeq
This model is called the modified Bonn quark model.\cite{A} 
In the previous paper, in Ref.\citen{A}, which is hereafter referred to as (A), 
exact eigenstates were investigated by the method of the boson realization. 
In Ref.\citen{B}, which is hereafter referred to as (B), 
the exact eigenstates with single-quark, quark-pair and quark-triplet structures were 
treated in a unified way. 
Further, in Ref.\citen{C}, which is referred to as (C), 
a phase diagram was given on the $\chi$-$N$ plane, where $N$ represents a quark number. 
However, in the series of previous papers (A)$\sim$(C), a color neutral quark-triplet state 
was only realized as a color-singlet state. 

In the first paper of the present series, namely in Ref.\citen{I}, which is hereafter referred to as (I), 
the exact eigenstates are constructed so as to satisfy a certain condition which gives a color-singlet state 
in average.
As a result, the color-singlet state is obtained in the color-symmetric form. 
In this paper, which is the second paper of the present series, the ground-state energy is 
reinvestigated under the condition giving a color-singlet state. 
In each region of the force strength $\chi$, the character of ground state is investigated, 
in which the quark-pairing state or quark-triplet state as a nucleon may be realized 
with a distinct value of a certain variable which is regarded as the order parameter 
of a phase transition. 
It is shown that there are two transition regions with different values of the order parameter. 

This paper is organized as follows:
In the next section, the basic scheme for searching the ground-state energy is given 
in the modified Bonn quark model. 
In \S 3, the condition for the ground state is investigated and in \S 4, 
the ground state is determined and the ground-state energy is derived in each area 
of the force strength $\chi$. 
In \S 5, numerical analysis is given. 
The last section is devoted to concluding remarks.

\section{Scheme for searching the ground-state energies}

\begin{figure}[t]
\begin{center}
\includegraphics[height=5.5cm]{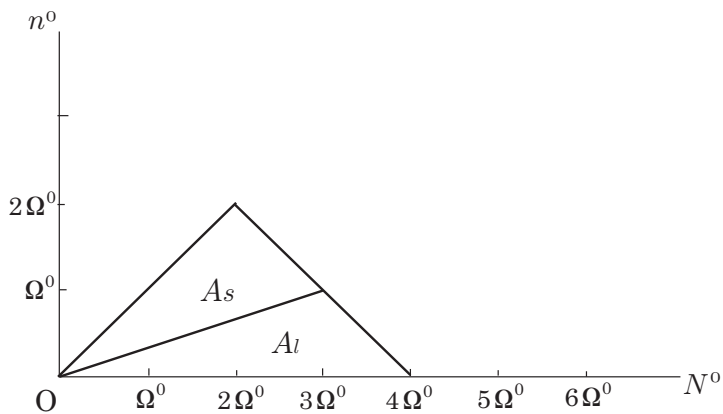}
\caption{Representation of the two areas of the $n^0$, $N^0$-plane considered in Eqs.(\ref{2-2a}) 
and (\ref{2-2b}).   
}
\label{fig:1}
\end{center}
\end{figure}

Our main concern in the present paper (II) is to search the minimum values 
of energies corresponding to the ground states. 
For the convenience of the discussion, we extract the relations 
(I$\cdot$6$\cdot$17) and (I$\cdot$6$\cdot$19): 
\beq\label{2-1}
E^{(m)}(N^0, n^0;2r)
&=&
\frac{1}{2}(1+2\chi)F(N^0,n^0;2r)+\frac{1}{2}n^0(2\Omega^0-n^0) \nonumber\\
& &-\frac{1}{6}N^0(6\Omega^0+6-N^0)\ , \qquad
(0\leq N^0 \leq 4\Omega^0) \qquad\qquad\ 
\eeq
\bsub\label{2-2}
\beq
E^{(m)}_l(N^0,n^0)
&=&E^{(m)}(N^0,n^0;2r=n^0)\nonumber\\
&=&\frac{1}{4}(1+6\chi)n^{02}-\frac{1}{2}\left[(N^0+3-2\Omega^0)+2\chi(N^0+3)\right]n^0
\nonumber\\
& &-\frac{1}{4}N^0(4\Omega^0+2-N^0)+\frac{\chi}{6}N^0(N^0+6) \quad
{\rm for\ the\ area}\ A_l \ , \quad
\label{2-2a}\\
E^{(m)}_s(N^0,n^0)
&=&E^{(m)}\left(N^0,n^0;2r=\frac{1}{2}(N^0-n^0)\right)\nonumber\\
&=&\frac{1}{4}(1+6\chi)n^{02}-\frac{1}{2}\left[(N^0-3-2\Omega^0)+2\chi(N^0-3)\right]n^0
\nonumber\\
& &-\frac{1}{4}N^0(4\Omega^0+6-N^0)+\frac{\chi}{6}N^0(N^0-6) \quad
{\rm for\ the\ area}\ A_s \ , \quad
\label{2-2b}
\eeq
\esub
Here, the areas $A_l$ and $A_s$ are shown in Fig.\ref{fig:1}. 
Later, we will treat $A_l$ by decomposing it into $A_{l_1}$ and $A_{l_2}$ which indicate the 
areas related to 
$0\leq N^0 \leq 3\Omega^0$ and $3\Omega^0 \leq N^0 \leq 4\Omega^0$, respectively. 
The relation (\ref{2-2}) shows us that $E_l^{(m)}(N^0,n^0)$ and 
$E_s^{(m)}(N^0,n^0)$ are quadratic in $n^0$ for a given value of $N^0$. 
In each area of $A_l$ and $A_s$, regarding $n^0$ as continuous variable, we first calculate 
the minimum values of $E_l^{(m)}(N^0,n^0)$ and $E_s^{(m)}(N^0,n^0)$, respectively, 
for a given value of $N^0$ and we compare with each other the two minima so obtained. 
The above is our scheme for searching the ground-state energies.

First, we notice that the term $F=(1/2)\cdot(1+2\chi)F(N^0,n^0;2r)$ in the 
expression (\ref{2-1}) is negative definite, if 
$\chi$ obeys 
\beq\label{2-3}
1+2\chi < 0\ , \quad {\rm i.e.,}\quad \chi < -\frac{1}{2} \ .
\eeq
Therefore, in this case, the maximum value of 
$F(N^0,n^0;2r)$ determines the minimum value of energy, which appears at 
the point $2r=0$, that is, $2s\neq 0$ or $2l \neq 0$. 
This case is contradictory to the requirement that 
$F(N^0,n^0;2r)$ should be as small as possible. 
Next, we consider the case 
\beq\label{2-4}
1+2\chi =0\ , \quad {\rm i.e.,}\quad \chi=-\frac{1}{2} \ .
\eeq
In this case, independently of the magnitude of $F(N^0,n^0;2r)$, the 
term $F$ vanishes. 
This indicates that the energies in the case $2r < 2r_m$ are the same as that in 
the case $2r=2r_m$, in other word, the case $2r=2r_m$ is not toward the smaller direction 
in energy. 
For the above reason, we will be concerned only with the case
\beq\label{2-5}
1+2\chi > 0 \ , \quad {\rm i.e.,}\quad 
\chi > -\frac{1}{2} \ .
\eeq
Certainly, the procedure for minimizing the energy in the case (\ref{2-5}) 
automatically leads to the condition required to $F(N^0,n^0;2r)$. 
The above tells that, in the case $\chi \leq -1/2$, the present 
model loses its meaning for the many-quark model. 
This point was already stressed in (A) qualitatively.

In (A) and (C), we showed the minimum value of the energy 
as a function of $N^0$ and the value of $n^0$ which minimizes the energy. 
In (A) and (C), we treated the case $n_0=0$, but essentially the same is valid for $n_0 \neq 0$. 
Also, we mentioned that $n^0$ can be regarded as the order parameter of the 
phase transition between the quark-pairs ($n^0=0$) and the quark-triplets ($n^0=N^0/3$). 
The results obtained in this investigation seem to be quite natural and 
acceptable. 
However, this investigation was not based on the ``color-singlet" states 
minimizing $F(N^0,n^0;2r)$. 
Further, the results were provided only for the area $A_l$. 
Therefore, we must reexamine the results presented in (A) and (C).

Following the procedure already mentioned, we are able to obtain the result in each area. 
In the area $A_l$, we obtain the following: 
\bsub\label{2-6}
\beq
& &(l_1;1)\quad -\frac{1}{2} < \chi \leq -\frac{1}{6}\cdot \frac{\Omega^0+6}{\Omega^0+2}\ , \nonumber\\
& &\qquad\quad {\rm (i)}\ \ n^0=0\quad {\rm for} \quad 
0 \leq N^0 \leq 3\Omega^0 \ , 
\label{2-6a}\\
& &(l_1;2)\quad -\frac{1}{6}\cdot \frac{\Omega^0+6}{\Omega^0+2} < \chi \leq -\frac{1}{6}\ , \nonumber\\
& &\qquad\quad {\rm (i)}\ \ n^0=0\quad {\rm for} \quad 
0 \leq N^0 \leq \frac{6(2\Omega^0-3-6\chi)}{5+6\chi} \ , \nonumber\\
& &\qquad\quad {\rm (ii)}\ \ n^0=\frac{N^0}{3} \quad {\rm for}\quad 
\frac{6(2\Omega^0-3-6\chi)}{5+6\chi} < N^0 \leq 3\Omega^0 \ ,  
\label{2-6b}\\
& &(l_1;3)\quad -\frac{1}{6} < \chi \leq \frac{2\Omega^0-3}{6}\ , \nonumber\\
& &\qquad\quad {\rm (i)}\ \ n^0=0\quad {\rm for} \quad 
0 \leq N^0 \leq \frac{2\Omega^0}{1+2\chi}-3 \ , \nonumber\\
& &\qquad\quad {\rm (ii)}\ \ n^0=\frac{(N^0-2\Omega^0+3)+2\chi(N^0+3)}{1+6\chi} \nonumber\\
& &\qquad\qquad\quad {\rm for}\quad 
\frac{2\Omega^0}{1+2\chi}-3 \leq N^0 \leq 3\Omega^0-\frac{9}{2}-9\chi \ , \nonumber\\
& &\qquad\quad {\rm (iii)}\ \ n^0=\frac{N^0}{3}\quad {\rm for} \quad 
3\Omega^0-\frac{9}{2}-9\chi \leq N^0 \leq 3\Omega^0 \ , 
\label{2-6c}\\
& &(l_1;4)\quad \frac{2\Omega^0-3}{6} < \chi < +\infty\ , \nonumber\\
& &\qquad\quad {\rm (i)}\ \ n^0=\frac{N^0}{3}\quad {\rm for} \quad 
0 \leq N^0 \leq 3\Omega^0 \ . 
\label{2-6d}
\eeq
\esub
The results (\ref{2-6}) coincide with those shown in the relation 
(C$\cdot$4$\cdot$1). 
In the case (\ref{2-6b}), we find that, at the point $N^0=6(2\Omega^0-3-6\chi)/(5+6\chi)$, 
a phase transition occurs. 
The order parameter $n^0$ changes from $n^0=0$ to 
$n^0=N^0/3$. 
In the area $A_{l_2}$, we obtain the following results: 
\bsub\label{2-7}
\beq
& &(l_2;1)\quad -\frac{1}{2} < \chi \leq -\frac{1}{2}\cdot \frac{2\Omega^0+3}{4\Omega^0+3}\ , \nonumber\\
& &\qquad\quad {\rm (i)}\ \ n^0=0\quad {\rm for} \quad 
3\Omega^0 \leq N^0 \leq 4\Omega^0 \ , 
\label{2-7a}\\
& &(l_2;2)\quad -\frac{1}{2}\cdot \frac{2\Omega^0+3}{4\Omega^0+3} < \chi \leq -\frac{1}{6}\ , \nonumber\\
& &\qquad\quad {\rm (i)}\ \ n^0=0\quad {\rm for} \quad 
3\Omega^0 \leq N^0 < \frac{2[(4\Omega^0-3)+6(2\Omega^0-1)\chi]}{3+10\chi} \ , \nonumber\\
& &\qquad\quad {\rm (ii)}\ \ n^0=4\Omega^0-N^0 \nonumber\\
& &\qquad\qquad\quad {\rm for}\quad 
\frac{2[(4\Omega^0-3)+6(2\Omega^0-1)\chi]}{3+10\chi} < N^0 \leq 4\Omega^0 \ ,  
\label{2-7b}\\
& &(l_2;3)\quad -\frac{1}{6} < \chi < +\infty \ , \nonumber\\
& &\qquad\quad {\rm (i)}\ \ n^0=4\Omega^0-N^0\quad {\rm for} \quad 
3\Omega^0 \leq N^0 \leq 4\Omega^0 \ .
\label{2-7c}
\eeq
\esub
In the case (\ref{2-7b}), the phase transition occurs at the point 
$N^0=2[(4\Omega^0-3)+6(2\Omega^0-1)\chi]/(3+10\chi)$. 
The parameter $n^0$ changes from $n^0=0$ to 
$n^0=4\Omega^0-N^0$. 
In the area $A_s$, we have the following: 
\bsub\label{2-8}
\beq
& &(s;1)\quad -\frac{1}{2} < \chi \leq -\frac{1}{6}\cdot \frac{4\Omega^0+9}{2\Omega^0+3}\ , \nonumber\\
& &\qquad\quad {\rm (i)}\ \ n^0=\frac{N^0}{3} \quad {\rm for} \quad 
0 \leq N^0 < \frac{3(2\Omega^0+3+6\chi)}{1-6\chi} \ , \nonumber\\
& &\qquad\quad {\rm (ii)}\ \ n^0=N^0 \quad {\rm for} \quad
\frac{3(2\Omega^0+3+6\chi)}{1-6\chi} < N^0 \leq 2\Omega^0 \ , 
\label{2-8a}\\
& &(s;2)\quad -\frac{1}{6}\cdot \frac{4\Omega^0+9}{2\Omega^0+3} < \chi < +\infty \ , \nonumber\\
& &\qquad\quad {\rm (i)}\ \ n^0=\frac{N^0}{3} \quad {\rm for} \quad 
0 \leq N^0 \leq 2\Omega^0 \ , 
\label{2-8b}\\
& &(s;3)\quad -\frac{1}{2} < \chi < +\infty \ , \nonumber\\
& &\qquad\quad {\rm (i)}\ \ n^0=\frac{N^0}{3} \quad {\rm for} \quad 
2\Omega^0 \leq N^0 \leq 3\Omega^0 \ .
\label{2-8c}
\eeq
\esub 
In the case (\ref{2-8a}), a phase transition occurs at the point 
$N^0=3(2\Omega^0+3+6\chi)/(1-6\chi)$. 
The parameter $n^0$ changes from 
$n^0=N^0/3$ to $n^0=N^0$.

As was shown in (A), the present model can be formulated not only from 
the side $N=0$ but also from the side $N=6\Omega$. 
We called this form the hole picture, which is obtained by 
replacing $N^0$ with $(6\Omega^0-N^0)$ in the relations 
appearing in the form from the side 
$N=0$. 
Therefore, it may be necessary to compare 
the results for $A_{l_2}$ with those for $A_{l_1'}$ obtained 
by replacing $N^0$ with $(6\Omega^0-N^0)$ in $A_{l_1}$. 
In the area $A_{l_1'}$, we obtain the following: 
\bsub\label{2-9}
\beq
& &(l'_1;1)\quad -\frac{1}{2} < \chi \leq -\frac{1}{6}\cdot \frac{\Omega^0+6}{\Omega^0+2}\ , \nonumber\\
& &\qquad\quad {\rm (i)}\ \ n'_0=0\quad {\rm for} \quad 
3\Omega^0 \leq N^0 \leq 6\Omega^0 \ , 
\label{2-9a}\\
& &(l'_1;2)\quad -\frac{1}{6}\cdot \frac{\Omega^0+6}{\Omega^0+2} < \chi \leq -\frac{1}{6}\ , \nonumber\\
& &\qquad\quad {\rm (i)}\ \ n'_0=2\Omega^0-\frac{N^0}{3} \quad {\rm for} \quad 
3\Omega^0 \leq N^0 < \frac{18(\Omega^0+1)(1+2\chi)}{5+6\chi} \ , \nonumber\\
& &\qquad\quad {\rm (ii)}\ \ n'_0=0 \quad {\rm for}\quad 
\frac{18(\Omega^0+1)(1+2\chi)}{5+6\chi} < N^0 \leq 6\Omega^0 \ ,  
\label{2-9b}\\
& &(l'_1;3)\quad -\frac{1}{6} < \chi \leq \frac{2\Omega^0-3}{6}\ , \nonumber\\
& &\qquad\quad {\rm (i)}\ \ n'_0=2\Omega^0-\frac{N^0}{3} \quad {\rm for} \quad 
3\Omega^0 \leq N^0 \leq 3\Omega^0+\frac{9}{2}+9\chi \ , \nonumber\\
& &\qquad\quad {\rm (ii)}\ \ n'_0=\frac{(4\Omega^0-N^0+3)+2\chi(6\Omega^0-N^0+3)}{1+6\chi} \nonumber\\
& &\qquad\qquad\quad {\rm for}\quad 
3\Omega^0+\frac{9}{2}+9\chi \leq N^0 \leq \frac{4(1+3\chi)\Omega^0}{1+2\chi}+3 \ , \nonumber\\
& &\qquad\quad {\rm (iii)}\ \ n'_0=0 \quad {\rm for} \quad 
\frac{4(1+3\chi)\Omega^0}{1+2\chi}+3 \leq N^0 \leq 6\Omega^0 \ , 
\label{2-9c}\\
& &(l'_1;4)\quad \frac{2\Omega^0-3}{6} < \chi < +\infty\ , \nonumber\\
& &\qquad\quad {\rm (i)}\ \ n'_0=2\Omega^0-\frac{N^0}{3} \quad {\rm for} \quad 
3\Omega^0 \leq N^0 \leq 6\Omega^0 \ . 
\label{2-9d}
\eeq
\esub

\section{Condition for the ground-state energy}

In \S 2, we listed the condition for minimizing the energy in each area. 
In order to search the ground-state energy, we must compare the minimum energy 
value in each area with the other areas. 
First, we discuss the relation between $A_{l_1}$ and $A_s$. 
As can be seen in the area $A_s$, the relation $n^0=N^0$ appears in the 
case $(s;1)$ (ii):
\beq\label{3-1}
-\frac{1}{2} < \chi \leq -\frac{1}{6}\cdot \frac{4\Omega^0+9}{2\Omega^0+3} \ , \quad
n^0=N^0 \quad {\rm for}\quad 
\frac{3(2\Omega^0+3+6\chi)}{1-6\chi} < N^0 \leq 2\Omega^0 \ . \quad
\eeq
In any other case, we have $n^0=N^0/3$. 
On the other hand, we notice the case $(l_1;1)$. 
Since $-(1/6)\cdot(4\Omega^0+9)/(2\Omega^0+3) < -(1/6)\cdot(\Omega^0+6)/(\Omega^0+2)$ 
and $2\Omega^0 < 3\Omega^0$, $E_s^{(m)}(N^0,n^0=N^0)$ should be 
compared with $E_l^{(m)}(N^0,n^0=0)$. 
For the comparison, we have the relation 
\beq\label{3-2}
\Delta E&=&
E_s^{(m)}(N^0,n^0=N^0)-E_l^{(m)}(N^0,n^0=0) \nonumber\\
&=& \frac{1}{2}(2\Omega^0-N^0)+\frac{1}{2}(1+2\chi)(N^0+2) \ .
\eeq
The relation (\ref{3-2}) indicates that, in the case (\ref{3-1}), 
$\Delta E$ is positive. 
In any other case in $A_s$, we have $n^0=N^0/3$ which 
belongs to the border of $A_s$ and $A_l$ and we notice the relation 
\beq\label{3-3}
E_s^{(m)}(N^0,n^0=N^0/3)=E_{l_1}^{(m)}(N^0,n^0=N^0/3)\ .
\eeq
Therefore, the comparisons of $E_s^{(m)}(N^0,n^0=N^0/3)$ with 
$E_{l_1}^{(m)}(N^0,n^0)$ is reduced to those of 
$E_{l_1}^{(m)}(N^0,n^0=N^0/3)$ with 
$E_{l_1}^{(m)}(N^0,n^0)$: 
The results are reduced to those shown in the relations (\ref{2-6a})$\sim$ 
(\ref{2-6d}). 
There does not exist any contribution of the area $A_s$ to the 
ground-state energy.

Next, we consider the overlap of the areas $A_{l_2}$ and $A_{l_1'}$. 
The overlap in this case consists of the regions composed of the 
following combinations: 
\bsub\label{3-4}
\beq
& &C_1\ ; \ \ n^0=0\ , \quad n^{0}{}'=0 \ , 
\label{3-4a}\\
& &C_2\ ; \ \ n^0=4\Omega^0-N^0\ , \quad n^{0}{}'=0 \ , 
\label{3-4b}\\
& &C_3\ ; \ \ n^0=0\ , \quad n^{0}{}'=2\Omega^0-\frac{N^0}{3} \ , 
\label{3-4c}\\
& &C_4\ ; \ \ n^0=4\Omega^0-N^0\ , \quad n^{0}{}'=2\Omega^0-\frac{N^0}{3} \ , 
\qquad\qquad\qquad\qquad\qquad\qquad
\label{3-4d}
\eeq
\esub
\beq\label{3-5}
& &C_5\ ; \ \ n^0=4\Omega^0-N^0\ , \quad n^{0}{}'=
\frac{(4\Omega^0-N^0+3)+2\chi(6\Omega^0-N^0+3)}{1+6\chi} \ . 
\eeq
For the combinations $C_1\sim C_4$, we can prove that $\Delta E 
(=E_{l_2}-E_{l_1'})$ are positive: 
\beq\label{3-6}
\Delta E > 0 \quad {\rm for}\quad C_1 \sim C_4 \ .
\eeq
For this proof, the following forms are useful: 
\bsub\label{3-7}
\beq
& &\Delta E=(1+2\chi)(\Omega^0+1)X \qquad\qquad\qquad\qquad\quad {\rm for}\quad C_1 \ , 
\label{3-7a}\\
& &\Delta E=\frac{1}{4}(3+10\chi)X^2+\frac{1}{2}(1+2\chi)(\Omega^0+5)X \nonumber\\
& &\qquad\quad
-\frac{1}{4}\left[(1+6\chi)\Omega^0+2(1+2\chi)\right]\Omega^0 
\qquad\quad {\rm for}\quad C_2 \ , 
\label{3-7b}\\
& &\Delta E=\frac{1}{36}(5+6\chi)X^2+\frac{1}{2}(1+2\chi)(\Omega^0+1)X \nonumber\\
& &\qquad\quad
+\frac{1}{4}\left[(1+6\chi)\Omega^0+6(1+2\chi)\right]\Omega^0 
\qquad\quad {\rm for}\quad C_3 \ , 
\label{3-7c}\\
& &\Delta E=\frac{8}{9}(1+3\chi)X^2+2(1+2\chi)X 
\qquad\qquad\quad\ {\rm for}\quad C_4 \ . 
\label{3-7d}
\eeq
\esub
Here, $X$ denotes
\beq\label{3-8}
X=N^0-3\Omega^0 \ .
\eeq

\begin{figure}[t]
\begin{center}
\includegraphics[height=5.5cm]{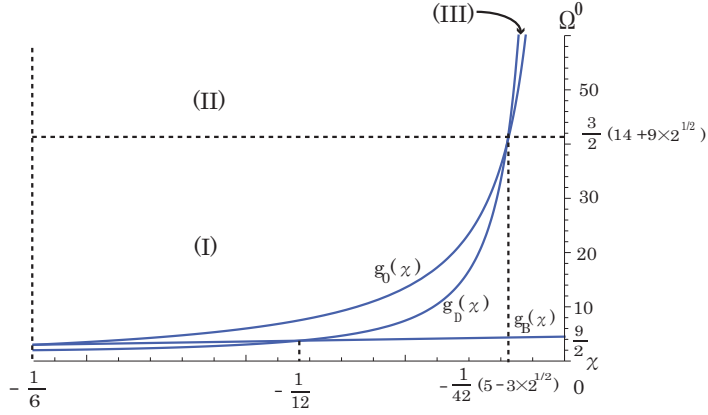}
\caption{The behavior of $g_B(\chi)$, $g_D(\chi)$ and $g_0(\chi)$ are illustrated.
In the area (I) and (II), $\Delta E <0$. The area (III) is divided into two areas. 
}
\label{fig:1-2}
\end{center}
\end{figure}

For the combination $C_5$, a rather lengthy consideration may be necessary. 
Our task is to investigate the overlap of the cases ($l_2$; 3(i)) and 
($l_1'$; 3(ii)) shown in the relations (\ref{2-7c}) and (\ref{2-9c}), respectively. 
In this case, the overlap range of $\chi$ is, at most, 
\beq\label{3-9}
-\frac{1}{6} < \chi < \frac{2\Omega^0-3}{6} \ .
\eeq
Further, we have the relation 
\beq\label{3-10}
3\Omega^0+\frac{9}{2}(1+2\chi) \leq N^0 \leq {\rm min}\left(\frac{4(1+3\chi)\Omega^0}{1+2\chi}+3,\ 4\Omega^0\right)\ .
\eeq
Of course, the relation (\ref{3-10}) gives us 
\beq\label{3-11}
3\Omega^0+\frac{9}{2}(1+2\chi) < 4\Omega^0\ .
\eeq
The relation (\ref{3-11}) is rewritten as 
\beq\label{3-12}
\Omega^0 > g_B(\chi)\ , \qquad g_B(\chi)=\frac{9}{2}(1+2\chi) \ , \quad {\rm i.e.,}\quad -\frac{1}{6} < \chi < \frac{2\Omega^0-9}{18} \ .
\eeq
The relation (\ref{3-10}) leads to the following relations:
\beq
& &g_B(\chi) < X < \frac{(\Omega^0+3)+6(\Omega^0+1)\chi}{1+2\chi} \ , 
\label{3-13}\\
& &g_B(\chi) < X < \Omega^0 \ .
\label{3-14}
\eeq
The relations (\ref{3-13}) and (\ref{3-14}) are valid, respectively in the ranges 
\beq
& &-\frac{1}{6} < \chi < -\frac{3}{2(2\Omega^0+3)} \ , 
\label{3-15}\\
& &-\frac{3}{2(2\Omega^0+3)} < \chi < \frac{2\Omega^0-9}{18} \ . 
\label{3-16}
\eeq
The energy difference $\Delta E$ is given as 
\beq\label{3-17}
\Delta E&=& \frac{4\chi(1+3\chi)}{1+6\chi}X^2+\frac{(1+2\chi)(1+12\chi)}{1+6\chi}X \nonumber\\
& &+\frac{6\chi(1+2\chi)}{1+6\chi}\Omega^0 + \frac{9}{4}\cdot\frac{(1+2\chi)^2}{1+6\chi}=f(X) \ .
\eeq
Since $X>0$, all the terms appearing in the function $f(X)$ are 
positive, if $\chi \geq 0$. 
Therefore, we have 
\beq\label{3-18}
{\rm if}\quad \chi \geq 0 \ , \quad
\Delta E > 0\quad{\rm in\ the\ range}\ \ X>0 \ .
\eeq
The above means that if $\chi \geq 0$, the energies in $A_{l_2}$ are larger than those of $A_{l_1'}$ for 
$N^0 > 3\Omega^0$. 
From the above reason, we will investigate $\chi < 0$ in the cases (\ref{3-13}) and (\ref{3-14}), 
together with the relations (\ref{3-15}) and (\ref{3-16}). 
Of course, the relation (\ref{3-16}) is changed to $-3/(2(2\Omega^0+3)) < \chi < 0$. 
Then, the relations (\ref{3-15}) and (\ref{3-16}) are, 
respectively, rewritten as 
\beq
& &\Omega^0 > g_0(\chi) \ , 
\label{3-19}\\
& &\Omega^0 < g_0(\chi) \ , 
\label{3-20}\\
& &g_0(\chi)=-\frac{3}{4}\cdot\frac{1+2\chi}{\chi}\ . \qquad
\left( -\frac{1}{6}< \chi < 0 \right)
\label{3-21}
\eeq

\begin{figure}[t]
\begin{center}
\includegraphics[height=7.0cm]{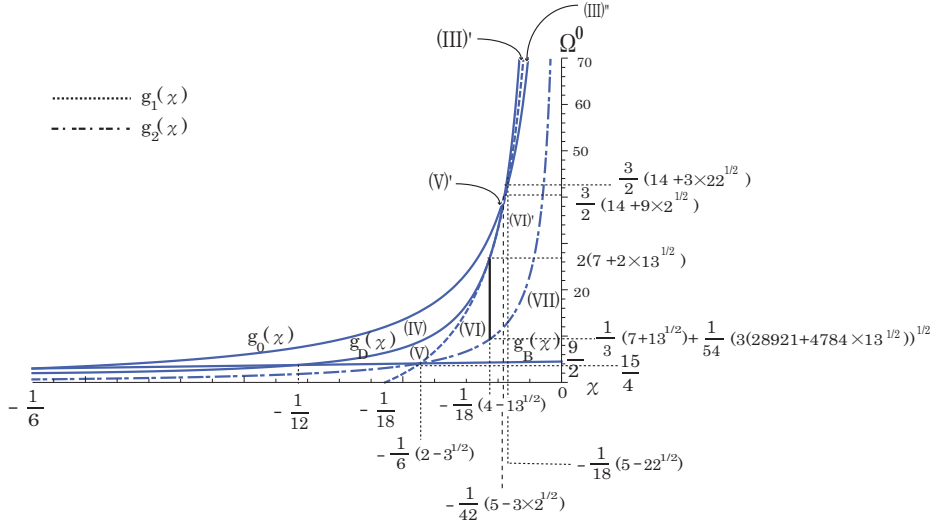}
\caption{
The behavior of $g_1(\chi)$ and $g_2(\chi)$ are illustrated together with $g_B(\chi)$, $g_D(\chi)$ and $g_0(\chi)$.
The area (III) is divided into two areas, (III)' and (III)". 
In the area (III)', $\Delta E <0$ and in (III)", the regions of $\Delta E <0$ and $\Delta E >0$ coexist. 
In the area (IV), $\Delta E <0$. In the area (V), $\Delta E <0$ and in (V)', the regions of 
$\Delta E<0$ and $\Delta E >0$ coexist. Also, in the area (VI) and (VI)', the regions of 
$\Delta E<0$ and $\Delta E >0$ coexist. In the area (VII), $\Delta E >0$. 
The area (V)' is too small to show the size definitely. 
}
\label{fig:2-2}
\end{center}
\end{figure}

First, we treat the case (\ref{3-13}) combined with the relations 
(\ref{3-12}) and (\ref{3-19}). 
The functions related to this case are $g_B(\chi)$, $g_0(\chi)$ and $g_D(\chi)$. 
The functions $g_B(\chi)$ and $g_0(\chi)$ are defined in the relations 
(\ref{3-12}) and (\ref{3-21}), respectively. 
The function $g_D(\chi)$ is defined through the discriminant of the 
quadratic function $f(X)$ given in the relation (\ref{3-17}), $D$:
\beq
& &D=-\frac{96\chi^2(1+2\chi)(1+3\chi)}{(1+6\chi)^2}\left(\Omega^0-g_D(\chi)\right)\ . 
\label{3-22}\\
& &g_D(\chi)=\frac{(1+2\chi)(1-6\chi)^2}{96\chi^2(1+3\chi)}\ .
\label{3-23}
\eeq
The behaviors of $g_B(\chi)$, $g_0(\chi)$ and $g_D(\chi)$ are illustrated in Fig.{\ref{fig:1-2}}. 
Since $g_B(\chi)>0$ and $g_0(\chi) >0$, $\Delta E <0$ and $\Delta E >0$ are determined 
by $\Omega^0 > g_D(\chi)$ and $\Omega^0 < g_D(\chi)$, respectively. 
In the areas (I) and (II), $\Delta E < 0$. 
In (III), $D>0$, but, in the present frame, $\Delta E > 0$ or $<0$ cannot be determined.

Next, we treat the case (\ref{3-14}) combined with the relations 
(\ref{3-12}) and (\ref{3-20}). 
In addition to $g_B(\chi)$, $g_0(\chi)$ and $g_D(\chi)$, we introduce the 
functions $g_1(\chi)$ and $g_2(\chi)$ which are defined as 
follows:
\bsub\label{3-24}
\beq
& & f(g_B(\chi))=\frac{6\chi(1+2\chi)}{1+6\chi}\left(\Omega^0-g_1(\chi)\right) = f_1 \ , 
\label{3-14a}\\
& & g_1(\chi)=-\frac{9}{8\chi}(1+2\chi)^2(1+18\chi)\ , \qquad\qquad\qquad\qquad\qquad\qquad\qquad\qquad
\label{3-24b}
\eeq
\esub
\bsub\label{3-25}
\beq
& & f(\Omega^0)=\frac{4\chi(1+2\chi)}{1+6\chi}\left(\Omega^0-\frac{9(1+2\chi)^2}{16\chi(1+3\chi)}\cdot\frac{1}{g_2(\chi)}
\right)\left(\Omega^0-g_2(\chi)\right)=f_2 \ , 
\label{3-25a}\\
& & g_2(\chi)=-\frac{1+2\chi}{8\chi(1+3\chi)}\left(1+18\chi+\sqrt{1+216\chi^2}\right) \ . 
\label{3-25b}
\eeq
\esub
Here, it should be noted that $(\Omega^0-9(1+2\chi)^2/(16\chi(1+3\chi))\cdot(1/g_2(\chi)))$ 
is positive. 
It may be clear from the definition of $g_i(\chi)$ for $i=1$ and 2, that if
$\Omega^0 < g_i(\chi)$, $f_i>0$ and vice versa.

\begin{figure}[t]
\begin{center}
\includegraphics[height=7.5cm]{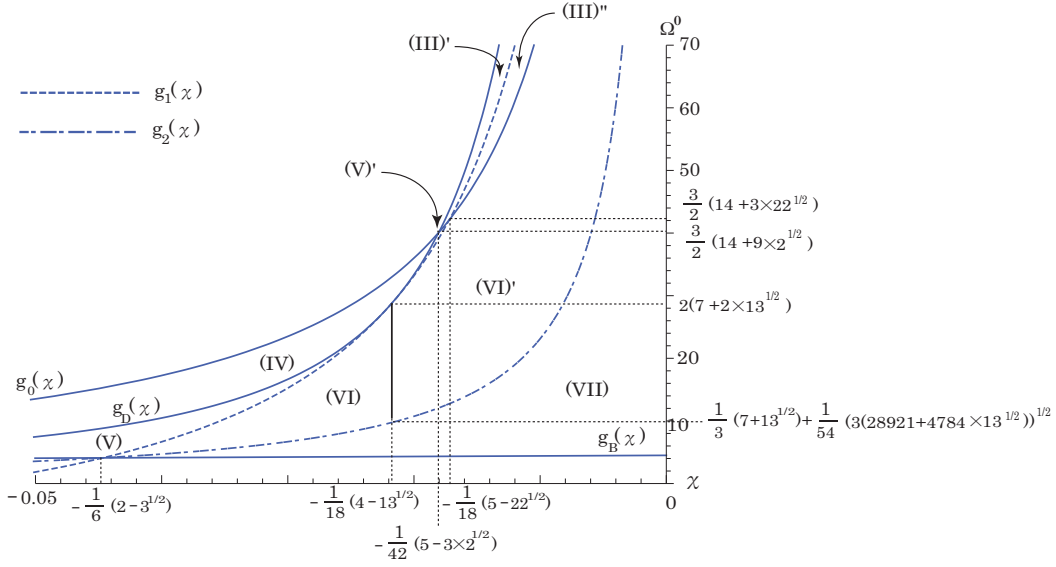}
\caption{
The detail around the area (V)' in Fig.3 is revealed. }
\label{fig:3-2}
\end{center}
\end{figure}

Figure {\ref{fig:2-2}} shows the behaviors of $g_1(\chi)$ and $g_2(\chi)$, together with 
$g_B(\chi)$, $g_0(\chi)$ and $g_D(\chi)$ which are shown in Fig.{\ref{fig:1-2}}. 
With the use of the relation between $g_i(\chi)$ and $f_i$, we obtain 
the following feature for $\Delta E<0$ and $\Delta E >0$: 
(i) In the area (IV), $\Delta E < 0$. 
(ii) In the area (V), $\Delta E < 0$ and in (V)', the 
regions of $\Delta E <0$ and $\Delta E > 0$ coexist. 
(iii) In the areas (VI) and (VI)', the regions of $\Delta E <0$ and $\Delta E>0$ coexist. 
(iv) In the area (VII), $\Delta E > 0$. 
Detail discussion on the coexistence of 
$\Delta E <0$ and $\Delta E>0$ is done in next section.

Finally, we discuss the area (III) in Fig.\ref{fig:1-2}. 
As is shown in Fig.\ref{fig:2-2}, the area (III) is divided into two areas, 
(III)' and (III)". 
In the area (III)', $\Delta E<0$ and in (III)", $\Delta E <0$ and $\Delta E>0$ coexist. 
In next section, we will discuss the meaning of the line 
$\chi=-(4-\sqrt{13})/18$.

\section{Determination of the ground-state energies}

We will continue the discussion in \S 3. 
The present model contains two parameters, $\Omega^0$ and $\chi$. 
The parameter $\Omega^0$ denotes the available degeneracy of the 
single-particle levels and determines the framework of the model. 
On the other hand, $\chi$ determines the dynamics caused by the model. 
Therefore, it may be natural to investigate the dynamics by 
changing $\chi$ for a fixed value of $\Omega^0$. 
From Figs.{\ref{fig:1-2}} and \ref{fig:2-2}, we can learn that if $\chi$ changes from 
$\chi=-1/6$ to $\chi=0$ for a given $\Omega^0$, there appear various features. 
The case $9/2 < \Omega^0 < \Omega_c^0=(7+\sqrt{13})/3+\sqrt{3(28921+4784\sqrt{13})}/54
(=10.42 \cdots)$ 
may be the simplest, but physically interesting. 
Other cases are trivial or too complicated. 
In this paper, we will investigate mainly this case.

Originally, $\Omega^0$ is a positive integer and in the present case, 
$\Omega^0$ takes the values as 
\beq\label{4-1}
& &\Omega^0=5,\ 6,\ 7,\ 8,\ 9\ {\rm or}\ 10.
\eeq
In (A) and (C), we showed various numerical results in the case $\Omega^0=6$. 
Therefore, the comparison of the present results with them may be interesting. 
We can classify the range $-1/6 < \chi < 0$ into the following cases:
\bsub\label{4-2}
\beq
& &(1)\ \ -\frac{1}{6}<\chi < \chi_0(\Omega^0)\quad {\rm in\ \ (I)} \ , 
\label{4-2a}\\
& &(2)\ \ \chi_0(\Omega^0) < \chi < \chi_1(\Omega^0)\quad {\rm in\ \ (IV)}+{\rm (V)} \ , 
\label{4-2b}\\
& &(3)\ \ \chi_1(\Omega^0) < \chi < \chi_2(\Omega^0)\quad {\rm in\ \ (VI)} \ , 
\label{4-2c}\\
& &(4)\ \ \chi_2(\Omega^0) < \chi < 0\quad {\rm in\ \ (VII)} \ . 
\label{4-2d}
\eeq
\esub
Here, $\chi_i(\Omega^0)$ $(i=0,\ 1,\ 2)$ denotes the inverse of $\Omega^0=g_i(\chi)$. 
The function $\chi_0(\Omega^0)$ is simply expressed as 
\bsub\label{4-3}
\beq
& &\chi_0(\Omega^0)=-\frac{3}{2(2\Omega^0+3)}\ . 
\label{4-3a}
\eeq
The function $\chi_1(\Omega^0)$ is obtained in an extremely complicated form: 
\beq\label{4-3b}
& &\chi_1(\Omega^0)=\left[
\sqrt[3]{-b+\sqrt{R}}+\sqrt[3]{-b-\sqrt{R}}-\frac{2}{27}Z^0\right]^{-1}\ , \nonumber\\
& &Z^0=4\Omega^0+99\ , \quad R=a^3+b^2\ , \nonumber\\
& &a=-\frac{4}{729}\left(Z^0{}^2-4617\right) \ , \quad 
b=\frac{4}{19683}\left(2Z^0{}^3-13851 Z^0+177147\right)\ . 
\eeq
The function $\chi_2(\Omega^0)$ is expressed in the form 
\beq\label{4-3c}
& &\chi_2(\Omega^0)=-\frac{1}{2}\cdot\frac{4\Omega^0+9}{
(2\Omega^0+1)(2\Omega^0+9)+2\Omega^0\sqrt{4\Omega^0{}^2+28\Omega^0+55}}\ . 
\eeq
For the case $\Omega^0=6$, we have 
\beq\label{4-3d}
\chi_0(6)=-0.1\ , \qquad
\chi_1(6)=-0.041099 \ , \qquad
\chi_2(6)=-0.032811 \ . 
\eeq
\esub

With the aid of the properties of the quadratic function $f(X)$ (in $-\infty < X < +\infty$), 
we have the following features:
\bsub\label{4-4}
\beq
& &(1)'\ \Delta E < 0\ \ {\rm for}\ \ g_B(\chi) < X < \frac{(\Omega^0+3)+6(\Omega^0+1)\chi}{1+2\chi} \ , 
\label{4-4a}\\
& &(2)'\ \Delta E < 0\ \ {\rm for}\ \ g_B(\chi) < X < \Omega^0 \ , 
\label{4-4b}\\
& &(3)'\ \Delta E > 0\ \ {\rm for}\ \ g_B(\chi) < X < X_L(\chi;\Omega^0) \ , \nonumber\\
& &\ \ \ \ \ \ \Delta E < 0\ \ {\rm for}\ \ X_L(\chi;\Omega^0) < X < \Omega^0\ , 
\label{4-4c}\\
& &(4)'\ \Delta E > 0\ \ {\rm for}\ \ g_B(\chi) < X < \Omega^0 \ . 
\label{4-4d}
\eeq
\esub
Of course, $(1)' \sim (4)'$ correspond to $(1)\sim (4)$, respectively. 
Here, $X_L$ denotes the larger for two solutions of $f(X)=0$ and 
it is explicitly given as 
\bsub\label{4-5}
\beq\label{4-5a}
X=-\frac{(1+2\chi)(1+12\chi)}{8\chi(1+3\chi)}
+\sqrt{\frac{3}{2}}\sqrt{\frac{1+2\chi}{1+3\chi}}
\sqrt{g_D(\chi)-\Omega^0}
=X_L(\chi;\Omega^0)\ . 
\eeq
Here, $f(X)$ is given in the relation (\ref{3-17}). 
The equation $f(X)=0$ is also regarded as quadratic for $\chi$ and the larger solution is 
obtained in the form 
\beq\label{4-5b}
\chi&=&
-\frac{1}{2}\cdot\frac{4X+9}{2(3\Omega^0+X(2X+5))+(4X+9)+2\sqrt{(3\Omega^0+X(2X+5))^2
-X^2(4X+9)}} \nonumber\\
&=&\chi_L(X;\Omega^0) \ .
\eeq
\esub
The first term in the relation (\ref{4-5a}) denotes the value of $X$ which makes $f(X)$ 
maximum and it is smaller than $g_B(\chi)$ in the case 
\beq\label{4-6}
\chi < -\frac{4-\sqrt{13}}{18}\ .
\eeq
Therefore, the smaller solution loses its meaning.

Until the present, we have investigated the effect coming from the area $A_{l_2}$ in the 
range $3\Omega^0 \leq N^0 \leq 4\Omega^0$. 
By replacing $N$ with $(6\Omega-N)$, the above effect can be included in the 
range $2\Omega^0 \leq N^0 \leq 3\Omega^0$. 
Including other areas, we can arrange the results in the ground-states as follows: 
\bsub\label{4-7}
\beq
& &(1) \ \ -\frac{1}{2} < \chi < -\frac{1}{6}\cdot\frac{\Omega^0+6}{\Omega^0+2}\ , \nonumber\\
& &\ \ \ \ {\rm (i)}\ \ 0 \leq N^0 \leq 3\Omega^0 , \qquad n^0=0 \ , 
\label{4-7a-a}\\
& &(2) \ \ -\frac{1}{6}\cdot\frac{\Omega^0+6}{\Omega^0+2} < \chi < -\frac{1}{6}\ , \nonumber\\
& &\ \ \ \ {\rm (i)}\ \ 0 \leq N^0 < \frac{6(2\Omega^0-3-6\chi)}{5+6\chi}\ , \qquad n^0=0 \ , \nonumber\\
& &\ \ \ \ {\rm (ii)}\ \ \frac{6(2\Omega^0-4-6\chi)}{5+6\chi} < N^0 \leq 3\Omega^0\ , \qquad n^0=\frac{N^0}{3} \ ,
\label{4-7a-b}\\
& &(3) \ \ -\frac{1}{6} < \chi < \chi_0(\Omega^0)\ , \nonumber\\
& &\ \ \ \ {\rm (i)}\ \ 0 \leq N^0 < \frac{2\Omega^0}{1+2\chi}-3\ , \qquad n^0=0 \ , \nonumber\\
& &\ \ \ \ {\rm (ii)}\ \ \frac{2\Omega^0}{1+2\chi}-3 < N^0 < 3\Omega^0-\frac{9}{2}(1+2\chi)\ , \qquad n^0=n^0{}^{\dagger} \ , \nonumber\\
& &\ \ \ \ {\rm (iii)}\ \ 3\Omega^0-\frac{9}{2}(1+2\chi) < N^0 \leq 3\Omega^0 \ , \qquad n^0=\frac{N^0}{3} \ , 
\label{4-7a}\\
& &(4) \ \ \chi_0(\Omega^0) < \chi < \chi_1(\Omega^0)\ , \nonumber\\
& &\ \ \ \ {\rm (i)}\ \ 0 \leq N^0 < \frac{2\Omega^0}{1+2\chi}-3\ , \qquad n^0=0 \ , \nonumber\\
& &\ \ \ \ {\rm (ii)}\ \ \frac{2\Omega^0}{1+2\chi}-3 < N^0 < 2\Omega^0\ , \qquad n^0=n^0{}^* \ , \nonumber\\
& &\ \ \ \ {\rm (iii)}\ \ 2\Omega^0 < N^0 < 3\Omega^0-\frac{9}{2}(1+2\chi) \ , \qquad n^0=n^0{}^{\dagger} \ , \nonumber\\
& &\ \ \ \ {\rm (iv)}\ \ 3\Omega^0-\frac{9}{2}(1+2\chi) < N^0 \leq 3\Omega^0 \ , \qquad n^0=\frac{N^0}{3} \ , 
\label{4-7b}\\
& &(5) \ \ \chi_1(\Omega^0) < \chi < \chi_2(\Omega^0)\ , \nonumber\\
& &\ \ \ \ {\rm (i)}\ \ 0 \leq N^0 < \frac{2\Omega^0}{1+2\chi}-3\ , \qquad n^0=0 \ , \nonumber\\
& &\ \ \ \ {\rm (ii)}\ \ \frac{2\Omega^0}{1+2\chi}-3 < N^0 < 2\Omega^0\ , \qquad n^0=n^0{}^* \ , \nonumber\\
& &\ \ \ \ {\rm (iii)}\ \ 2\Omega^0 < N^0 < 3\Omega^0-X_L(\chi;\Omega^0) \ , \qquad n^0=n^0{}^{\dagger} \ , \nonumber\\
& &\ \ \ \ {\rm (iv)}\ \ 3\Omega^0 - X_L(\chi;\Omega^0) < N^0 < 3\Omega^0-\frac{9}{2}(1+2\chi) \ , \qquad n^0=n^0{}^* \ , \nonumber\\
& &\ \ \ \ {\rm (v)}\ \ 3\Omega^0-\frac{9}{2}(1+2\chi) < N^0 \leq 3\Omega^0 \ , \qquad n^0=\frac{N^0}{3} \ , 
\label{4-7c}\\
& &(6) \ \ \chi_2(\Omega^0) < \chi < \frac{2\Omega^0-3}{6}\ , \nonumber\\
& &\ \ \ \ {\rm (i)}\ \ 0 \leq N^0 < \frac{2\Omega^0}{1+2\chi}-3\ , \qquad n^0=0 \ , \nonumber\\
& &\ \ \ \ {\rm (ii)}\ \ \frac{2\Omega^0}{1+2\chi}-3 < N^0 < 3\Omega^0-\frac{9}{2}(1+2\chi)\ , \qquad n^0=n^0{}^* \ , \nonumber\\
& &\ \ \ \ {\rm (iii)}\ \ 3\Omega^0-\frac{9}{2}(1+2\chi) < N^0 \leq 3\Omega^0 \ , \qquad n^0=\frac{N^0}{3} \ , 
\label{4-7d}\\
& &(7) \ \  \frac{2\Omega^0-3}{6} < \chi < +\infty\ , \nonumber\\
& &\ \ \ \ {\rm (i)}\ \ 0 \leq N^0 \leq 3\Omega^0\ , \qquad n^0=\frac{N^0}{3} \ .
\label{4-7e}
\eeq
\esub
Here, $n^0{}^{\dagger}$ and $n^0{}^*$ denote
\beq\label{4-8}
& &n^0{}^{\dagger}=N^0-2\Omega^0\ , \qquad
n^0{}^*=\frac{(N^0-2\Omega^0+3)+2\chi(N^0+3)}{1+6\chi}\ .
\eeq
If $n^0{}^{\dagger}$ is replaced with $n^0{}^*$, the result (\ref{4-7}) reduces to the 
result (C$\cdot$4$\cdot$1). 
With the use of the expression (\ref{2-2a}), the ground-state energies 
are calculated except the case $n^0=n^0{}^{\dagger}$, in which the energies 
are calculated by the relation (\ref{2-2a}) under the replacement of 
$N^0$ with $(6\Omega^0-N^0)$.

\section{Numerical analysis}

\begin{figure}[b]
\begin{center}
\includegraphics[height=4.5cm]{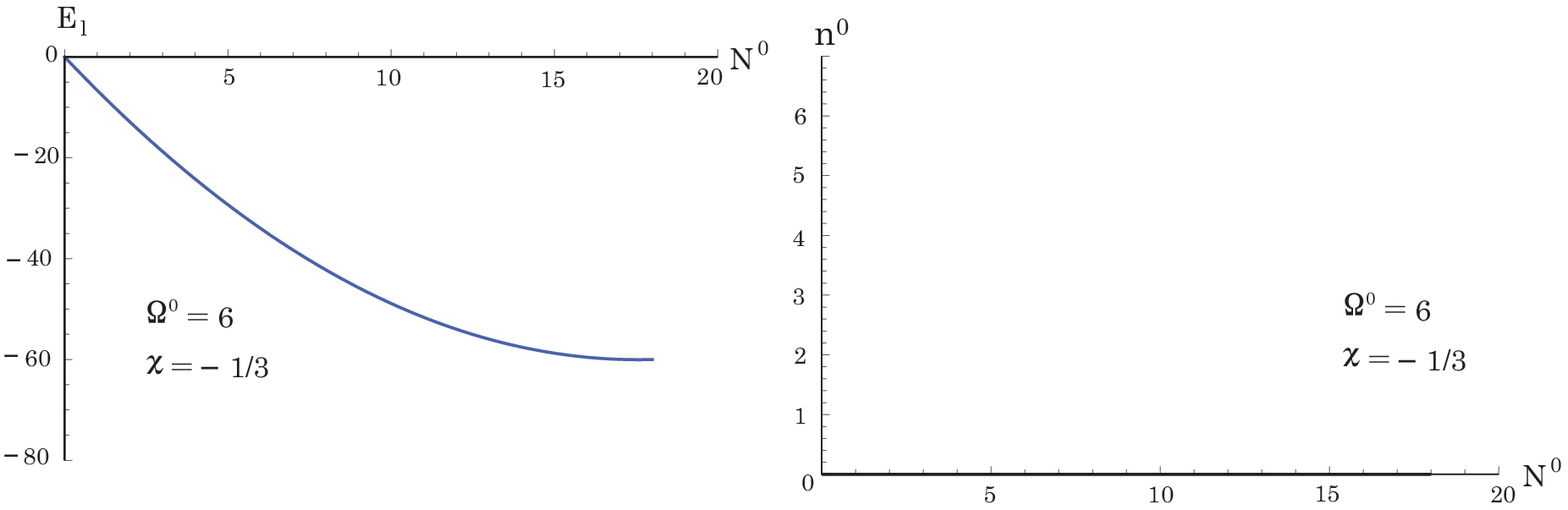}
\caption{The behaviors of energy (left panel) and the order parameter $n^0$ (right panel) 
are shown as functions of $N^0$ in the case (1) in Eq.(\ref{4-7a-a}) with $\chi=-1/3$. 
The model parameter $\Omega^0$ is taken as 6.
}
\label{fig:4-1}
\end{center}
\end{figure}
\begin{figure}[b]
\begin{center}
\includegraphics[height=4.5cm]{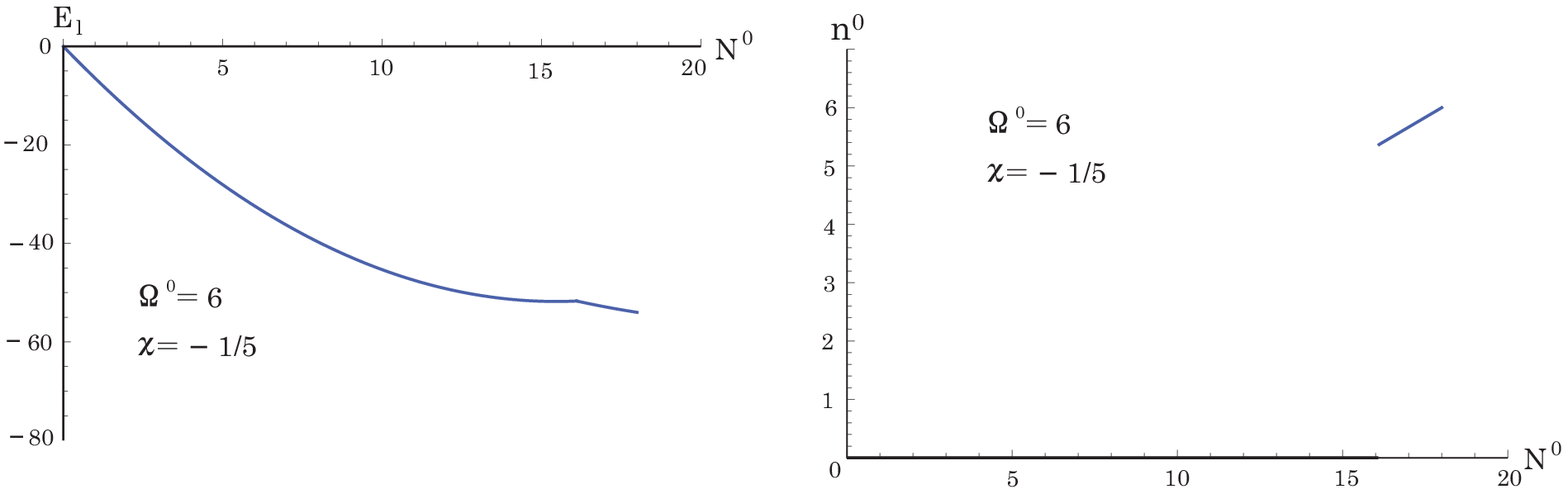}
\caption{The behaviors of energy (left panel) and the order parameter $n^0$ (right panel) 
are shown as functions of $N^0$ in the case (2) in Eq.(\ref{4-7a-b}) with $\chi=-1/5$. 
The model parameter $\Omega^0$ is taken as 6.
}
\label{fig:4-2}
\end{center}
\end{figure}

In this section, we give numerical results in the region $A_{l_1}$. 
The quantity $n^0$, which is regarded as an order parameter of a phase transition between 
the quark-triplets and the quark-pairs, 
is given in Eq.(\ref{4-7}) in the various areas. 
The ground-state energies are also calculated numerically by using Eq.(\ref{2-2a}). 
In this section, we fix the parameter $\Omega^0=6$. Thus, we show the behaviors of the 
ground-state energy and the order parameter in the region $0\leq N^0 \leq 3\Omega^0\ (=18)$ 
with various force strength of the particle-hole interaction, $\chi$. 

In Fig.\ref{fig:4-1}, the ground-state energy on the left panel and the order parameter on the right panel are,  
respectively, shown as a function of $N^0$ with $\chi=-1/3$, namely the results are based on Eq.(\ref{4-7a-a}). 
In this case, the order parameter $n^0$ is identical to 0. 
Thus, the quark-pair state is realized. 
Almost the same behavior is seen in Fig.2 in (A). 

In Fig.\ref{fig:4-2}, the ground-state energy and the order parameter are shown 
in the case $\chi=-1/5$, namely the results are based on Eq.(\ref{4-7a-b}). 
In this case, the order parameter $n^0$ changes from 0 to $N^0/3$. 
Thus, as the particle number $N^0$ increases, the quark-pair state changes to 
the quark-triplet state directly. 
This behavior is seen in Fig.3 in (A).

\begin{figure}[t]
\begin{center}
\includegraphics[height=4.5cm]{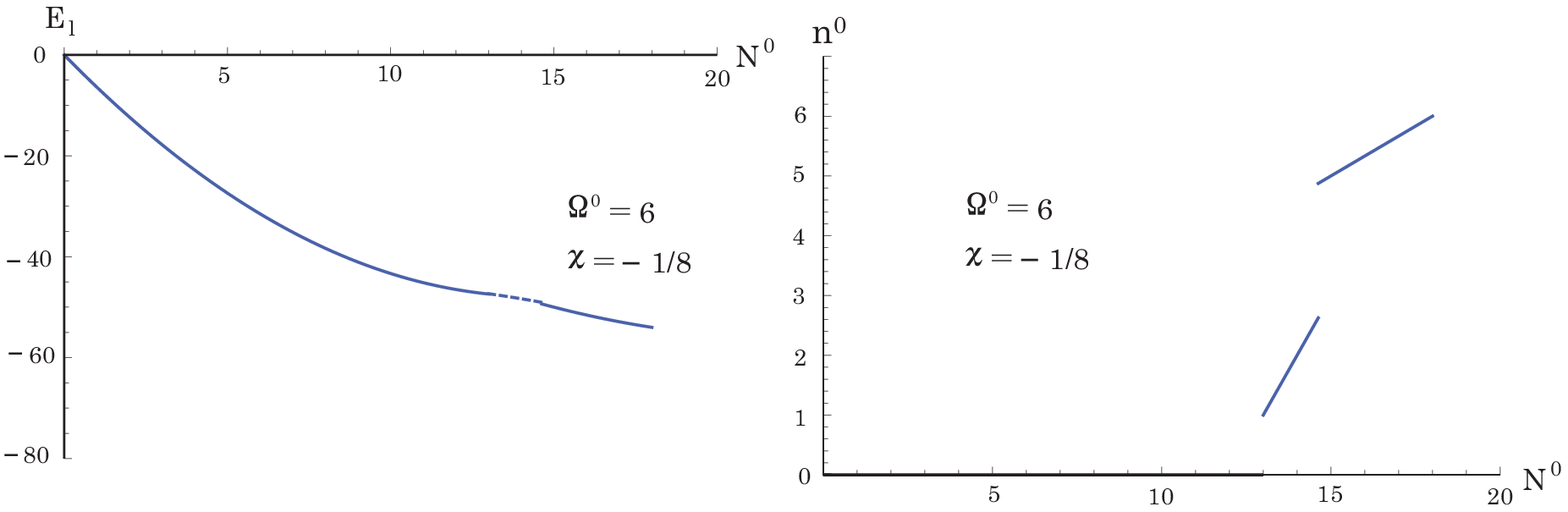}
\caption{The behaviors of energy (left panel) and the order parameter $n^0$ (right panel) 
are shown as functions of $N^0$ in the case (3) in Eq.(\ref{4-7a}) with $\chi=-1/8$. 
The model parameter $\Omega^0$ is taken as 6.
}
\label{fig:4-3}
\end{center}
\end{figure}
\begin{figure}[t]
\begin{center}
\includegraphics[height=4.5cm]{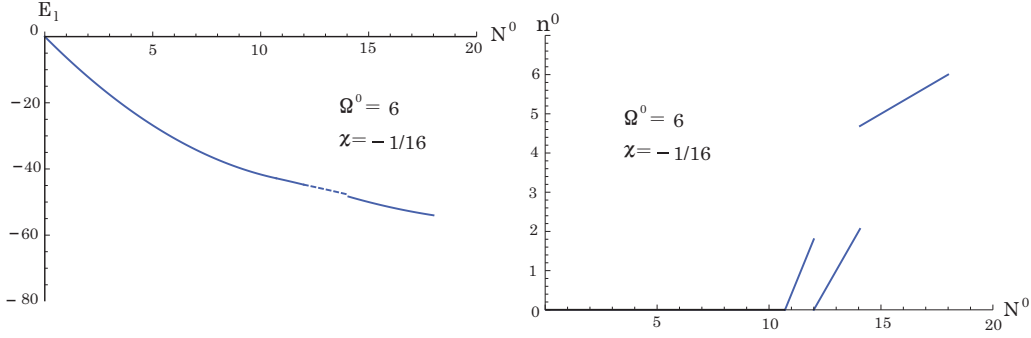}
\caption{The behaviors of energy (left panel) and the order parameter $n^0$ (right panel) 
are shown as functions of $N^0$ in the case (4) in Eq.(\ref{4-7b}) with $\chi=-1/16$. 
The model parameter $\Omega^0$ is taken as 6.
}
\label{fig:4-4}
\end{center}
\end{figure}

However, in Fig.\ref{fig:4-3}, another behavior of the order parameter $n^0$ is seen. 
The ground-state energy and the order parameter are shown 
in the case $\chi=-1/8$ based on Eq.(\ref{4-7a}). 
In this case, the order parameter $n^0$ changes from 0 to $N^0/3$. 
However, in the region $13 < N^0 < 14.625$ (if $N^0$ is integer, $13 < N^0 \leq 14$), 
the transition region from quark-pair state to color-singlet quark-triplet state 
is open with $n^0=n^{0\dagger}$. 
The change of order parameter is not continuous, while the order parameter is 
changed continuously shown in Fig.4 in (A).
For the point between $n^0=0$ or $n^0=N^0/3$ and $n^0=n^{0\dagger}$, the ground-state energy is not continuously changed 
with respect to the change of $N^0$. 
The dotted curve in the left panel represents the ground state energy with $n^0=n^{0\dagger}$. 
However, as the particle number $N^0$ increases, the quark-pair state changes to 
the quark-triplet state through the transition region.

\begin{figure}[t]
\begin{center}
\includegraphics[height=4.5cm]{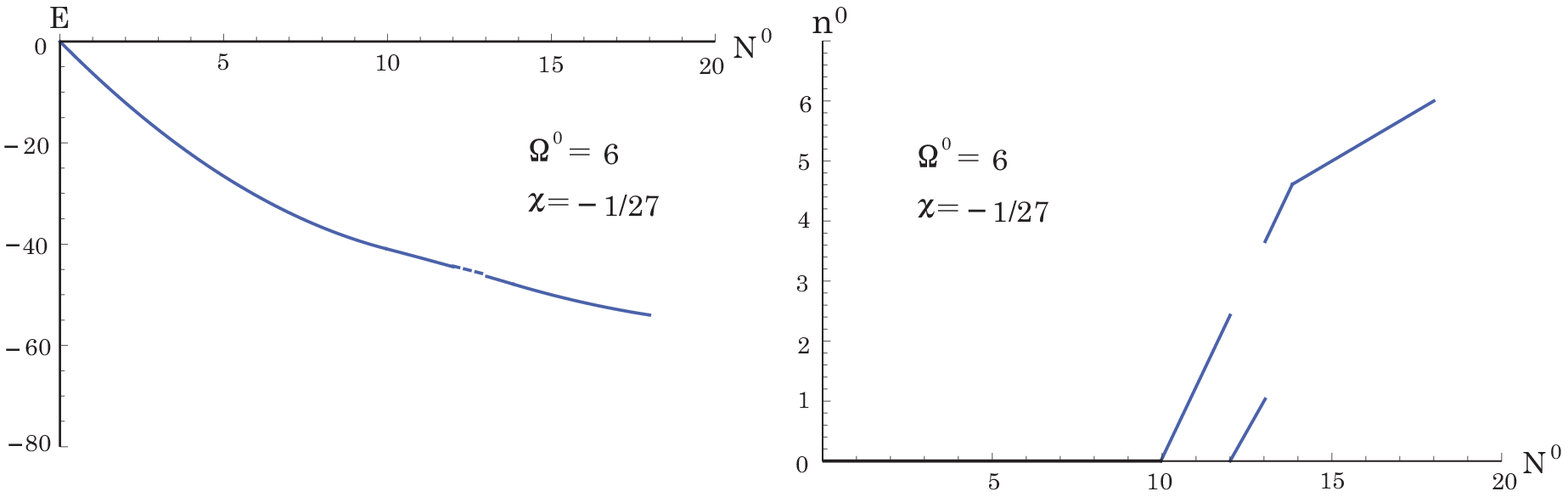}
\caption{The behaviors of energy (left panel) and the order parameter $n^0$ (right panel) 
are shown as functions of $N^0$ in the case (5) in Eq.(\ref{4-7c}) with $\chi=-1/27$. 
The model parameter $\Omega^0$ is taken as 6.
}
\label{fig:4-5}
\end{center}
\end{figure}
\begin{figure}[t]
\begin{center}
\includegraphics[height=4.5cm]{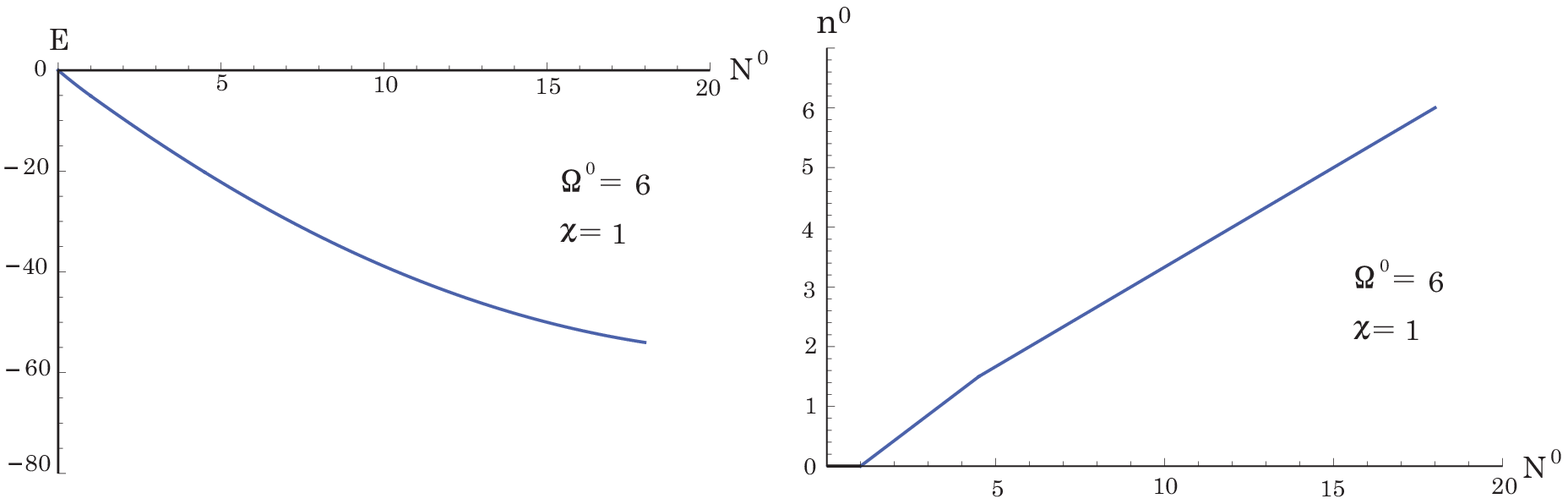}
\caption{The behaviors of energy (left panel) and the order parameter $n^0$ (right panel) 
are shown as functions of $N^0$ in the case (5) in Eq.(\ref{4-7d}) with $\chi=1$. 
The model parameter $\Omega^0$ is taken as 6.
}
\label{fig:4-6}
\end{center}
\end{figure}

In Fig.\ref{fig:4-4}, the ground-state energy and the order parameter are shown 
in the case $\chi=-1/16$ based on Eq.(\ref{4-7b}). 
In this case, the order parameter $n^0$ changes from 0 to $N^0/3$ through 
$n^0=n^{0*}$ and $n^0=n^{0\dagger}$. 
For the point between $n^0=n^{0*}$ or $n^0=N^0/3$ and $n^0=n^{0\dagger}$, the ground-state energy is not continuously changed 
with respect to the change of $N^0$. 
The dotted curve in the left panel represents the ground state energy with $n^0=n^{0\dagger}$. 
In this case, also, as the particle number $N^0$ increases, the quark-pair state changes to 
the quark-triplet state through two transition regions characterized by $n^{0*}$ and $n^{0\dagger}$. 

Also, in Fig.\ref{fig:4-5} in the case $\chi=-1/27$ based on Eq.(\ref{4-7c}), 
the order parameter $n^0$ changes from 0 to $N^0/3$ through 
$n^0=n^{0*}$, $n^0=n^{0\dagger}$ and $n^0=n^{0*}$. 
For the point between $n^0=n^{0*}$ and $n^0=n^{0\dagger}$, the ground-state energy is not 
continuously changed with respect to the change of $N^0$ similar to the case $\chi=-1/8$.  
Here, the dotted curve in the left panel represents the ground state energy with $n^0=n^{0\dagger}$. 
In this case, also, as the particle number $N^0$ increases, the quark-pair state changes to 
the quark-triplet state through three transition regions.

\begin{figure}[t]
\begin{center}
\includegraphics[height=4.5cm]{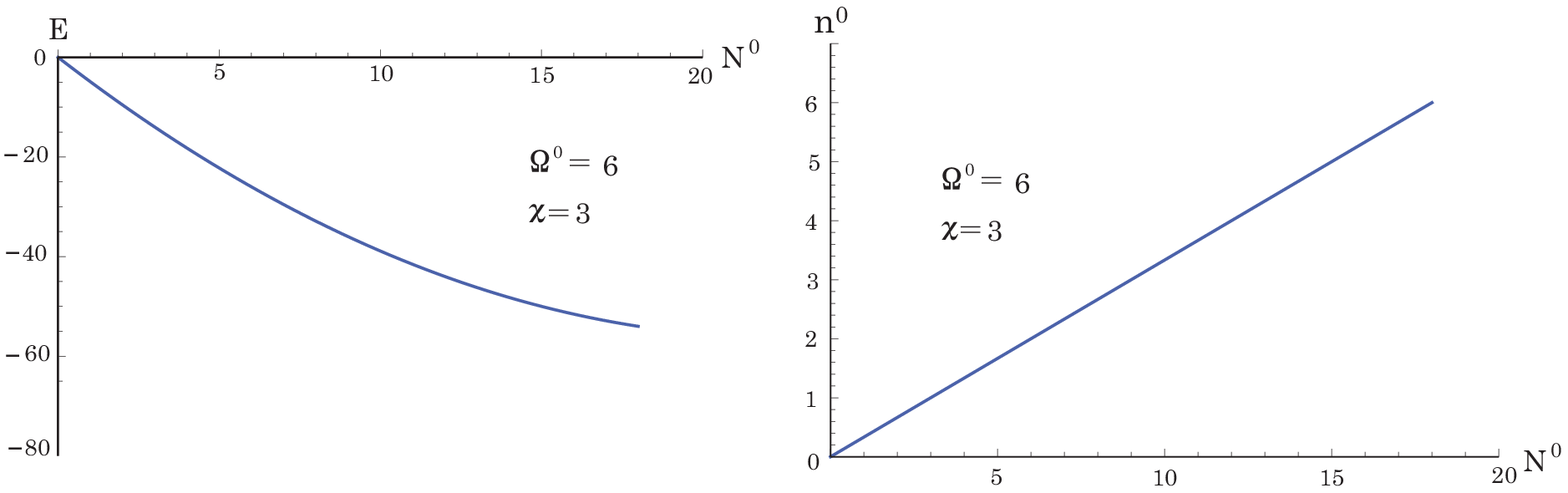}
\caption{The behaviors of energy (left panel) and the order parameter $n^0$ (right panel) 
are shown as functions of $N^0$ in the case (5) in Eq.(\ref{4-7e}) with $\chi=3$. 
The model parameter $\Omega^0$ is taken as 6.
}
\label{fig:4-7}
\end{center}
\end{figure}

In Fig.\ref{fig:4-6}, the ground-state energy and the order parameter are shown 
in the case $\chi=1$, namely the results are based on Eq.(\ref{4-7d}). 
In this case, the order parameter $n^0$ changes from 0 to $N^0/3$ through the 
transition region with $n^0=n^{0*}$. 
Thus, as the particle number $N^0$ increases, the quark-pair state changes to 
the quark-triplet state through the transition region. 
This behavior corresponds to Fig.5 in (A).

Finally, in Fig.\ref{fig:4-7}, the ground-state energy and the order parameter are shown 
in the case $\chi=3$, namely the results are based on Eq.(\ref{4-7e}). 
In this case, the order parameter $n^0$ is identical to $N^0/3$ in all the regions.  
Thus, the color-singlet quark-triplet state is realized.  
This behavior of the order parameter corresponds to that shown in Fig.6 in (A).

\section{Concluding remarks}

\begin{figure}[t]
\begin{center}
\includegraphics[height=7.5cm]{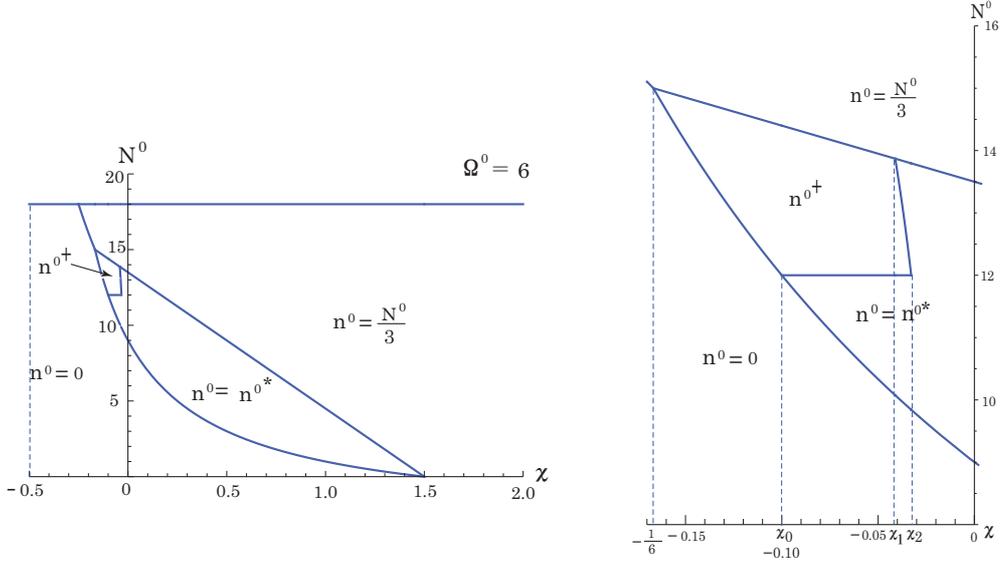}
\caption{The phase diagram is shown in the $\chi$-$N^0$ plane (left panel). The order parameter 
is regarded as $n^0$. In the right panel, the details are shown in the region of $-0.17 < \chi < 0$. 
Here, $\chi_i\ (i=0,\ 1, \ 2)$ represent $\chi_i(\Omega^0)$, respectively.}
\label{fig:5-1}
\end{center}
\end{figure}

In this paper, the ground-state energies were calculated in the modified Bonn quark model. 
The color-singlet condition was imposed, which was developed in the previous paper.\cite{I} 
It was indicated that the modified Bonn quark model has a meaning in the regions
$\chi>-1/2$, where $\chi$ represents the strength of the particle-hole-type interaction. 
The ground-state energies were determined in each area of $\chi$ and the 
ground states were also determined. 
Namely, in a certain parameter region, the color-singlet quark-triplet state is the 
ground state, and in another region, the quark-color-pairing state is the 
ground state. 
Further, it was shown that there were two transition regions between the quark-pairing and 
the quark-triplet state, which were distinguished by the value of the 
order parameter.

Finally, we give a comment on the phase transition 
induced by the present model. 
For this aim, we show the phase diagram obtained by the relation (\ref{4-7}). 
It is drawn for the case $\Omega^0=6$ in Fig.{\ref{fig:5-1}}. 
Since $n^0$ plays a role of the order parameter, the present model 
induces four phases and the areas specified by $n^0=N^0/3$ and 
$n^0=0$ are in the quark-triplet and the quark-pair phase, respectively. 
In the original Bonn model, the case $\chi=0$ was adopted. 
The region $N^0 \sim 0$ or $6\Omega^0$ and the region 
$N^0\sim 3\Omega^0$ are in the quark-pair and the quark-triplet phase, 
respectively. 
Further, we must note that in the former and the latter regions the quark system has a low and a 
high density, respectively. 
We observe, however, that the situation predicted by the present model is in contradiction 
with the common understanding,\cite{cs} according to which color-superconductivity is 
expected to provide a reliable description of quark matter at high densities. 
%
In the case where $\chi$ is 
different from $\chi=0$, the 
situation does not change, if we treat $\chi$ as a constant in the 
whole region $0\leq N^0 \leq 6\Omega^0$. 
Although the original and the modified Bonn quark model present us 
various features of many-quark system, it may be not 
permitted for these models to avoid the investigation of the 
above-mentioned problem. 
In (III),\cite{III} 
we will discuss this problem.

\end{document}